\documentclass[aip,jcp,reprint,superscriptaddress]{revtex4-1}
\usepackage{amsmath}
\usepackage{amssymb}
\usepackage{booktabs}
\usepackage{bm}
\usepackage{graphicx}
\usepackage[caption=false]{subfig}
\usepackage{dcolumn}
\usepackage{array}
\usepackage{hyperref}
\usepackage{orcidlink}

\begin{document}

\title{Polarisation-mediated underscreening from weakly bonded ion clusters}

\author{David Ribar}
\affiliation{Computational Chemistry, Lund University, P.O. Box 124, S-221 00 Lund, Sweden}

\author{Jake W. Felber}
\affiliation{Computational Chemistry, Lund University, P.O. Box 124, S-221 00 Lund, Sweden}
\affiliation{Heidelberg University, Faculty of Chemistry and Earth Sciences, Im Neuenheimer Feld 234, 69120 Heidelberg, Germany}

\author{Clifford E. Woodward}
\affiliation{School of Physical, Environmental and Mathematical Sciences University College, University of New South Wales, ADFA, Canberra, ACT 2600, Australia}

\author{Jan Forsman}
\email[]{jan.forsman@chem.lu.se}
\affiliation{Computational Chemistry, Lund University, P.O. Box 124, S-221 00 Lund, Sweden}

\date{\today}

\begin{abstract}
  We explore the hypothesis that ions form loosely connected clusters at high ionic strength in aqueous solutions, and that these clusters have relevance to the experimentally observed
  phenomenon usually referred to as ``anomalous underscreening''. Cluster formation lowers the ionic strength below its nominal value, slowing the decay of the screening length but not
  reversing it. Here we focus on an additional contribution, cluster polarisation. We demonstrate that this effect produces longer-ranged repulsive interactions between like-charged
  surfaces or particles in concentrated salt solutions. We derive an analytical bulk relation in which the entire architecture of a cluster enters through a single quantity, the
  charge-weighted second moment of its intramolecular charge structure factor, which quantifies the polarisation response of an arbitrary cluster topology. We also make numerical
  calculations using classical polymer Density Functional Theory, cDFT, for a model based on star-like clusters in which satellite ions are weakly bonded by a harmonic spring to
  a common central ion. The corona of satellite ions is assumed to be net neutral, since the formation of highly charged clusters would be accompanied by a
  significant self-energy cost. Using this model, we calculate surface interactions and screening lengths at various overall salt concentrations. We show that, under the
  assumption that the fraction of ions belonging to clusters increases with salt concentration, one may qualitatively arrive at ``anomalous underscreening'', i.e., an
  effective screening length that displays a minimum at an overall (monovalent) salt concentration of about $1\,\mathrm{M}$. Quantitatively, we note that the predicted
  growth of the screening length beyond this threshold value is weaker than typically found by experimental surface force measurements. 
\end{abstract}


\maketitle 

\section{Introduction}

Aqueous electrolytes are among the oldest and best-studied systems in liquid-state theory \cite{Israelachvili:91,Evans:94,Holm:01}, yet the asymptotic decay of their electrostatic
correlations at high concentration remains contested. The mean-field Debye–Hückel theory captures the dilute limit, in which the electrostatic potential decays
exponentially, with a Debye
length $\lambda_D \propto 1/\sqrt{c}$, where $c$ is the salt concentration. However, its assumptions of point-like ions, a uniform background permittivity, 
negligible ion–ion correlations, and a linearised exponential become increasingly inaccurate at high $c$, where $\lambda_D$ becomes comparable to molecular and ionic
dimensions \cite{HansenMcDonald, Israelachvili:91, attard_asymptotic_1993, LeotedeCarvalho1994, Kirkwood1936}.

Surprisingly, surface-force apparatus (SFA) measurements in concentrated electrolytes and ionic liquids have reported decay lengths that \textit{increase} with
concentration, in direct contradiction to the expected Debye scaling \cite{Smith:2016, lee_underscreening_2017, D3FD00042G, Gebbie:13, Gebbie:15}. This behaviour
defines the so-called \textit{anomalous underscreening} regime, which is theorised to be a universal property of highly concentrated charged
fluids \cite{lee_underscreening_2017, safran_scaling_2023}.

Besides SFA experiments, similar anomalous observations have been linked to this speculated hypersaline phenomenon\cite{elliott_known-unknowns_2024}, including
charged colloid re-entrance \cite{Yuan:22, Reinertsen2024}, polyelectrolyte brush re-entrant swelling \cite{Dunlop2026, Robertson2023}, and charged dye surface
excesses \cite{Gaddam2019}. With atomic force microscopy (AFM), there are published studies that directly contradict the existence of anomalous
underscreening\cite{Kumar:22}, while others report that the phenomenon is sensitive to instrumental parameters\cite{Tilger2026}. In a recent SFA
study of ionic liquids, \citet{Cross2026}  demonstrated the importance of long equilibration times to obtain a true equilibrium free-energy
measurement, although it should be noted that this was for rather viscous (non-aqueous) fluids. 

Rather than weigh the experimental evidence, we ask what mechanisms could underlie such behaviour, in aqueous solutions, where anomalous underscreening is
observed. This question
leads us to the structure of the bulk fluid at high salt concentrations, where there is published experimental evidence of clustering in aqueous
electrolytes\cite{Irving2023, Fetisov2020, Shalit2016, Straub2022, Sedlak:06a, Sedlak:06b, Gebauer2008, Demichelis2011}. Within
theoretical research, there are extensive molecular dynamics (MD) studies of highly concentrated NaCl that do not reproduce anomalous
underscreening \cite{Coles:20, Zeman:20}, while other studies with different molecular simulation force fields do identify ion clusters
that are hypothesised to be the reason for anomalous underscreening \cite{Krucker:21, Kim2014, Chen2007}. It should be noted that molecular
force fields are parametrised for a plethora of applications, with general force fields such as AMBER and OPLS succeeding in a wide variety
of simulations \cite{Cornell1995,Jorgensen1996}. These force fields may, however, be less reliable under extreme conditions, such as near-saturation
aqueous electrolytes, where the relevant configurations lie outside the parametrisation set \cite{Luo2009,Luo2013, Auffinger2007}.
Ion clusters and large screening lengths have been established, even within the restricted primitive model (RPM), albeit at electrostatic couplings much
stronger than in water \cite{Hartel:23}.

The ion-clustering hypothesis dates back to the original work of Bjerrum \cite{Bjerrum1926}, whereby ion pairing would deplete the system of
dissociated ions. In turn, this would render the effective concentration of ``free'' ions much lower than its nominal value, with a concomitant
increase of the screening length \cite{Gebbie:13, Gebbie:15}. However, it would be hard to reconcile this suggestion with the law of mass
action. As we add more salt to a solution, we would at least anticipate \textit{some} increase, and certainly not a decrease, in the concentration
of the dissociated species, even if the dissociated \textit{fraction} diminishes.

We have recently explored the hypothesis that ion clusters mediate surface interactions, by constructing such clusters explicitly\cite{Ribar:25a, Woodward2026, Ribar2026}. These
studies identified a strong correspondence between the dipolar nature of the polymeric cluster charge sequence and both the range and magnitude of the surface forces. In
this work, we isolate the underlying mechanism, using a different cluster model. While ion clusters
are expected to have a modest average net charge due to the self-energy cost of forming charged objects, they
nevertheless respond to external fields by \textit{polarising}. In fact, we demonstrate that cluster polarisation can generate strong additional repulsive contributions to
the interaction between like-charged surfaces, and hence an increase in the surface-force range, i.e., the effective screening length. 

\section{Model and Theory}
The main features of our model, as well as some useful analytic expressions, are presented below. The numerical methods, including a description of cDFT and the bulk simulation
methods, are
described in Sec. \ref{sec:methods}.

\subsection{The star cluster model}
Our model is based on the hypothesis that clusters with a rather ``loose'' structure and architecture (specifics given below), which we shall refer to
as \textit{star clusters}, are spontaneously formed in aqueous solutions at high ionic strengths. While similar phenomena possibly occur in ionic liquids and
their solvent mixtures, we focus here on simple models of water solutions.

What we propose is a gradual increase of ions belonging to clusters as salt is added, such that a \textit{fraction}, $\gamma$, of the added ions ends up in
clusters. We envisage that $\gamma$ increases with the overall salt concentration $c$, but that it saturates to a constant value at very high ionic strengths. Specifically, we assume that
\begin{equation}
\gamma(c) =
\begin{cases}
(0.9/2000^2)\,(c/\text{mM})^2 , & c < 2000\,\text{mM} \\[1ex]
0.9 , & c \geq 2000\,\text{mM} .
\end{cases}
\label{eq:gamma}
\end{equation}
By integrating $\gamma(c)$, we obtain the total concentration, $c_m$, of cations (or anions) that belong to (are monomers in) a star cluster as $c_m = \int_0^c \gamma(x)\,dx$, so that
\begin{equation}
c_m/\text{mM} =
\begin{cases}
(0.3/2000^2)\,(c/\text{mM})^3 , & c < 2000\,\text{mM} \\[1ex]
600 + 0.9\,(c/\text{mM}-2000) , & c \geq 2000\,\text{mM} .
\end{cases}
\label{eq:nm}
\end{equation}
The adopted values are of course to some extent arbitrary, but they find some support from the typical observation of a conductivity that  monotonically increases with
concentration, but displays some signs of saturation at ionic strengths above about $1\,\mathrm{M}$. 

Building up a substantial net charge on an ion cluster typically amounts to a significant free energy cost, and for the main part of this work, we assume that the star
clusters are monovalent, and thus effectively form a 1:1 salt, with a cluster salt concentration $c_c$. At each concentration the clusters are monodisperse, with $r$ denoting
the number of ions per cluster. However, $r$ does depend on $c$, and is adjusted so that $c_c$ approaches a plateau of about $30\,\mathrm{mM}$ for
high $c$. Table \ref{tab:cluster_concentrations} summarises the various concentrations and cluster sizes relevant to this work. We use $c_d$ to indicate
the concentration of dissociated salt, while $c_m$ is the total concentration of cationic (or anionic) ``monomers'' belonging to clusters. 

\begin{table}[htbp]
  \caption{Concentrations and cluster sizes resulting from the model assumptions. Here, $c$ is the total salt concentration, $c_m$ is the concentration of
    cations (or anions) belonging to clusters, $r$ is the number of ions per cluster, $c_c$ is the cluster salt concentration, and $c_d$ is the
    concentration of (free) dissociated salt. The listed values of $c_c$ and $c_d$ are rounded, but it should be noted that $c=c_d+c_m$, with $c_m=r\cdot c_c$.}
\begin{ruledtabular}
\begin{tabular}{ccccc}
$c$ (mM) & $c_m$ (mM) & $r$ & $c_c$ (mM) & $c_d$ (mM) \\
\hline
 100  &    0.075 &  3 &  0.025 &   99.925 \\
 200  &    0.6   &  3 &  0.2   &  199.4   \\
 400  &    4.8   &  3 &  1.6   &  395.2   \\
 600  &   16.2   &  3 &  5.4   &  583.8   \\
 800  &   38.4   &  3 & 12.8   &  761.6   \\
1000  &   75.0   &  3 & 25.0   &  925.0   \\
1200  &  129.6   &  5 & 25.9   & 1070.4   \\
1400  &  205.8   &  7 & 29.4   & 1194.2   \\
1600  &  307.2   & 11 & 27.9   & 1292.8   \\
1800  &  437.4   & 15 & 29.2   & 1362.6   \\
2000  &  600.0   & 21 & 28.6   & 1400.0   \\
2200  &  780.0   & 27 & 28.9   & 1420.0   \\
2400  &  960.0   & 33 & 29.1   & 1440.0   \\
2800  & 1320.0   & 45 & 29.3   & 1480.0   \\
3200  & 1680.0   & 57 & 29.5   & 1520.0   \\
3600  & 2040.0   & 69 & 29.6   & 1560.0   \\
4000  & 2400.0   & 81 & 29.6   & 1600.0   \\
\end{tabular}
\end{ruledtabular}
\label{tab:cluster_concentrations}
\end{table}

\begin{figure}
	\centering
	\includegraphics[scale=0.4]{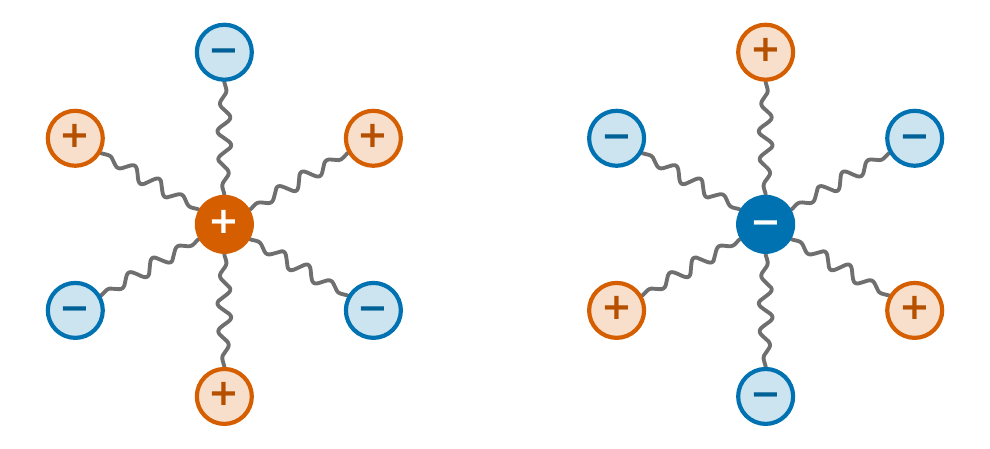}
	\caption{An illustration of a 7-mer ($r=7$) star cluster salt, composed of a cationic star cluster (left) and an anionic star cluster (right). Note
          that the central monomers dictate the valency of each cluster, as the $r-1$ satellite monomers are charge-neutral in total ($(r-1)/2$ satellite
          cations and $(r-1)/2$ satellite anions). The satellites are drawn evenly spaced for clarity, and no particular arrangement is implied. The
          squiggly lines represent the loose harmonic bonds of Eq. (\ref{eq:hbond}); these are likewise drawn with equal lengths, whereas the actual
          bond lengths are distributed according to the Boltzmann weight of that equation.}
	\label{fig:cartoon}
\end{figure}

An important aspect of the clusters is that they are held together rather ``loosely'', which we regard as physically reasonable. If the solution were to form tight
solid-like clusters of significant size, they would most likely manifest themselves experimentally by very strong scattering peaks. As far as we are aware, such
strong and distinct peaks have never been found. On the other hand, some experiments do report rather broad and weak scattering peaks for multivalent
ions\cite{Dinpajooh2024,Reinertsen2024}, which are at least qualitatively consistent with our model. We describe a star cluster composed of $r$ monomers, each
of which is an ion, consisting of a central monomer to which $r-1$ satellite monomers are connected via rather ``sloppy'' harmonic bonds, creating a satellite
ion corona. Of the satellites, $(r-1)/2$ are cationic and $(r-1)/2$ anionic, so that they carry no net charge. Consequently, the number of monomers in a cluster
is odd, and the valency of a cluster is dictated by its central monomer. The model is illustrated in Figure \ref{fig:cartoon}.
The harmonic bond potential, $V_b$, is given by 
\begin{equation}
  \beta V_b = f|{\bf r'}-{\bf r}|^2 ,
    \label{eq:hbond}    
\end{equation}
where $\beta = 1/(k_BT)$ is the inverse thermal energy, while ${\bf r'}$ and ${\bf r}$ are the coordinates of a satellite and a central ion. The bond
constant, $f$, regulates the tightness of the cluster.

All species (dissociated as well as clustered ions) are charged and carry a common hard-sphere diameter $d = 3\,\text{\AA}$. Furthermore, the solvent is treated
implicitly by setting a uniform dielectric constant of $\varepsilon_r = 78.3$, corresponding to that of water at $T = 298\,\text{K}$. The corresponding Bjerrum
length is $l_B=7.16\,\text{\AA}$. Bond stiffnesses are expressed in terms of the reduced constant $K^*\equiv fl_B^2$. We
consider $K^*=0.856$ and $1.712$ for the main part of this work. We use $\eta$ to indicate the valency of an individual ion, i.e., $\eta=\pm1$, and $Q$ for
the net-valency of a cluster. Non-bonded interactions between all species are described by the pair potential $\beta\phi_{ij}(s)$:
\begin{equation}
\beta\phi_{ij}(s) =
\begin{cases}
\infty , & s \leq d \\[1ex]
l_B\,\dfrac{\eta_i \eta_j}{s} , & s > d ,
\end{cases}
\label{eq:pairpot}
\end{equation}
where $s$ is the separation between ions $i$ and $j$.

\subsection{Analytical model and the intramolecular charge structure factor}
\label{sec:theory:omega}

Seeking an analytical model for cluster screening, we simplify the ion cluster model to a point-charge star. A central monovalent ion at index $i=0$ is surrounded
by a \emph{corona} of $r-1$ satellites, indices $i=1,\dots,r-1$, each tethered to the core by the harmonic bond of Eq.~\eqref{eq:hbond}. Satellites are statistically
independent and all bonds share the same stiffness $f$. The harmonic tether makes each satellite a classical Drude oscillator. Each site carries a physical charge
of $\eta_ie$. The corona is electroneutral, $\sum_{i\ge1}\eta_i=0$, so the cluster carries the valency of its core, $Q\equiv\eta_0+\sum_{i\ge1}\eta_i=\pm1$. In this
point-charge idealisation, the resulting bond-length distribution is Gaussian with width $\xi^2=1/(2f)$. The hard-core interactions and intermolecular
correlations of the full model are not included in this analytical approximation. The following section is an encapsulation of the theoretical results, with the
derivations reserved for Appendix \ref{app:first}.

The object that carries all of the cluster's internal electrostatics is its \emph{intramolecular charge structure factor},
\begin{equation}
    \omega_c(k) =  \sum_{i=0}^{r-1}\sum_{j=0}^{r-1}\eta_i\eta_j
    \big\langle e^{\mathrm{i} \mathbf{k}\cdot(\mathbf{r}_i-\mathbf{r}_j)}\big\rangle ,
    \label{eq:omega_def}
\end{equation}
the charge-weighted analogue of the single-molecule form factor familiar from  scattering theory\cite{HansenMcDonald}. Note that the upright
symbol $\mathrm{i}$ denotes the imaginary unit, $\mathrm{i}=\sqrt{-1}$, whereas the italic $i$ is a site index. Only relative positions
enter, so $\omega_c$ is independent of cluster position and of any choice of cluster centre. Expanding at small $k$, the odd orders vanish identically, yielding
\begin{equation}
    \omega_c(k) = Q^{2} + \tfrac{1}{3}(r-1)R_c^{2}\,k^{2} + \mathcal{O}(k^{4}).
    \label{eq:omega_smallk}
\end{equation}
We then define the charge-weighted electrostatic spread per satellite, $R_c$, through
\begin{equation}
    R_c^{2} \equiv \frac{3}{r-1}\lim_{k\to0}
    \frac{\omega_c(k)-Q^{2}}{k^{2}} .
    \label{eq:rc_def}
\end{equation}
Because $R_c$ is built from $\omega_c$ rather than from a moment about a chosen point, it is automatically translation invariant and may be sampled directly
from raw simulation coordinates via
\begin{equation}
    (r-1)R_c^{2} = \big\langle|\mathbf{P}|^{2}\big\rangle
    - Q\sum_{i=0}^{r-1}\eta_i\big\langle|\mathbf{r}_i|^{2}\big\rangle,
    \label{eq:C2split}
\end{equation}
where $\mathbf{P} \equiv \sum_i\eta_i\mathbf{r}_i$ is the instantaneous reduced dipole moment (the physical dipole
being $e\mathbf{P}$), and $\chi_c$ (defined in Eq.~\eqref{eq:chi_c} below) therefore obtainable for any cluster architecture, including
those for which $\omega_c(k)$ has no closed form.

For the harmonic star the double sum in Eq.~\eqref{eq:omega_def} closes in elementary form,
\begin{equation}
    \omega_c(k) = r - (r-1)e^{-k^{2}\xi^{2}},
    \label{eq:omega_star}
\end{equation}
with $R_c^{2} = 3\xi^{2} = 3/(2f)$, so that $R_c$ is the root-mean-square bond length for this architecture. Its values interpolate
between $\omega_c(0)=Q^{2}$ and $\omega_c(\infty)=r$. The identification $R_c^{2}=3/(2f)$ holds because of corona electroneutrality
and uniform bond stiffness making the second term on the r.h.s. of Eq.~\eqref{eq:C2split} vanish. For a charge-asymmetric corona, or
for non-uniform bond stiffness, one or the other condition fails, in which case Eq.~\eqref{eq:C2split} must be retained in full.

The route taken here differs from mean-field treatments of ion association, in which the cluster enters through a dipole moment $\mu=be$ whose
separation $b$ is supplied as a model parameter \cite{Adar2017, Zwanikken2009}. Retaining $\omega_c(k)$ instead determines the dipole from the
bonding architecture and stiffness.

\subsection{Linear response and the screening pole}
\label{sec:theory:pole}

Point ions have no internal structure, $\omega_{\rm ion}(k)=1$, and are present
at number density $n_d$ of each sign, while clusters contribute $\omega_c(k)$ at
number density $n_c$. The bare charge correlation per unit volume is therefore
\begin{equation}
    \omega_{\rm tot}(k) = 2n_d + 2n_c\,\omega_c(k).
    \label{eq:omegatot}
\end{equation}
A cluster responds to an external potential as a unit rather than site by site. Linearising its Boltzmann weight and letting each modulated cluster deposit charge
at all of its sites returns precisely Eq.~\eqref{eq:omega_def}, giving $\delta\tilde\rho_q(k) = -\beta e^{2}\omega_{\rm tot}(k)\tilde\psi(k)$. With Poisson's
equation and $\beta e^{2}/\varepsilon_0\varepsilon_r = 4\pi l_B$,
\begin{equation}
    \tilde\psi(k) = \frac{q_0}{\varepsilon_0\varepsilon_r}\,
    \frac{1}{k^{2}+4\pi l_B\,\omega_{\rm tot}(k)} .
    \label{eq:psi}
\end{equation}
The asymptotic decay of the real-space potential is governed by the pole of
Eq.~\eqref{eq:psi} nearest the real axis, i.e., by the root of
\begin{equation}
    k^{2} + 4\pi l_B\,\omega_{\rm tot}(k) = 0
    \label{eq:pole}
\end{equation}
with smallest imaginary part \cite{Kjellander1992, attard_asymptotic_1993,
  Evans1994, LeotedeCarvalho1994}. The imaginary part sets the effective screening length, $\lambda_{\rm eff}\equiv1/\mathrm{Im}\,k$. A purely imaginary
root $k=\mathrm{i}\kappa_{\rm eff}$ gives monotonic Yukawa decay, while a conjugate pair gives damped oscillatory decay. We restrict attention to parameter
regimes in which the nearest pole is the purely imaginary one.

\subsection{Master screening equation}
\label{sec:theory:master}

Truncating $\omega_c$ at its second moment, Eqs.~\eqref{eq:omega_smallk} and \eqref{eq:omegatot} reduce Eq.~\eqref{eq:pole} to an algebraic equation
with a purely imaginary root. With the satellite screening constant $\kappa_{\rm sat}^{2}\equiv8\pi l_B(r-1)n_c$,
\begin{equation}
    \kappa_{\rm eff}^{2} = \frac{8\pi l_B(n_d+n_cQ^2)}{1+\chi_c},
    \label{eq:master}
\end{equation}
and 
\begin{equation}
    \chi_c \equiv \tfrac{1}{3}\kappa_{\rm sat}^{2}R_c^{2}.
    \label{eq:chi_c}
\end{equation}
Referencing to the Debye length of the dissociated salt alone,
$\lambda_D\equiv(8\pi l_B n_d)^{-1/2}$,
\begin{equation}
    \frac{\lambda_{\rm eff}}{\lambda_D} = \sqrt{1+\chi_c}\sqrt{C},
    \label{eq:ratio}
\end{equation}
with the concentration correction factor
\begin{equation}
    C \equiv \frac{n_d}{n_d+n_cQ^2}.
\end{equation}
For the monovalent clusters considered here, $Q^2=1$.

Equation~\eqref{eq:ratio} is a Debye--Langevin susceptibility in disguise.
Writing $$\chi_c=\tfrac13\kappa_{\rm sat}^{2}R_c^{2}=n_{\rm sat}\alpha_{\rm bond}/\varepsilon_0\varepsilon_r$$ where $n_{\rm sat}=2(r-1)n_c$ is the total satellite
number density and $\alpha_{\rm bond}=\tfrac13\beta e^{2}R_c^{2}$, identifies $R_c^{2}$ as a polarisability per bonded monomer. The static susceptibility
follows from the zero-field mean-square dipole by the standard fluctuation formula \cite{Frohlich1958}. The resulting competition between dielectric
enhancement and free-ion depletion parallels the mean-field Bjerrum-pair treatment
of \citet{Adar2017}, whose $\varepsilon_{\rm eff}=\varepsilon+ \tfrac13\beta n_p p^{2}$ and $\lambda_{\rm eff}=(\varepsilon_{\rm eff}/2\beta e^{2}n_s)^{1/2}$ are
the counterparts of our $\chi_c$ and $C$.

For the harmonic star, Eq.~\eqref{eq:omega_star} gives $R_c^{2}=3/(2f)$, so that
\begin{equation}
    \chi_c = \frac{4\pi l_B(r-1)n_c}{f} = \frac{4\pi l_B^3(r-1)n_c}{K^{*}},
    \label{eq:chiAK}
\end{equation}
utilizing $K^{*} \equiv f l_B^{2}$ as the reduced bond stiffness measured at the Bjerrum length. It is convenient to rearrange Eq.~\eqref{eq:ratio} as
\begin{equation}
    \frac{\lambda_{\rm eff}^2}{\lambda_D^2C} -1= \chi_c.
    \label{eq:theory_prediction}
\end{equation}
Because $\chi_c\propto1/K^*$ for ideal Gaussian bonds, $K^{*}$ acts as a polarisability-softness dial, with soft bonds allowing large thermal satellite
deformations and thus a highly polarisable cluster, while stiff bonds a nearly inert one. We further identify  $\chi_c\propto n_c$ and $\chi_c\propto (r-1)$, with
higher cluster concentrations and more bonded satellites producing a stronger polarisation response. 

We also note that it is possible to extend the present second-order theory to an exact solution by inserting a closed-form
expression for the intramolecular charge structure factor $\omega_c$ directly into Eq.~\eqref{eq:pole}. The resulting
transcendental equation can be readily solved with standard numerical techniques to obtain the exact effective screening
length in the context of the present theory. Concretely, for our charge architecture, the equation reads as
\begin{equation}
	\kappa_{\rm eff}^{2} = 8\pi l_B\left[n_d + n_c\left(r-(r-1)
	e^{\kappa_{\rm eff}^{2}\xi^{2}}\right)\right],
	\label{eq:exact_pole}
\end{equation}
where the exponential arises from continuing Eq.~\eqref{eq:omega_star} to $k=i\kappa_{\rm eff}$. The same construction
applies to any architecture for which $\omega_c$ closes in elementary form, such as the rigid zwitterionic dimer treated in
Appendix~\ref{app:first}.

\section{Methods}
\label{sec:methods}
\subsection{Classical density functional theory}
The calculations are performed adopting the numerical implementation that is standard to polymer-cDFT. Hard-sphere exclusion between particles is approximately
accounted for, and Coulomb interactions are managed at the mean-field level. We consider a system consisting of two infinite, rigid, and planar
surfaces, immersed in the model mixture, and separated by a distance $h$. Each surface carries a uniform charge density, $\sigma_s =  -0.014286 \,e/${\AA}$^2$, which is
close to estimated values for the mica surfaces that are utilised in SFA experiments \cite{Crothers:21}.

The system is described within the grand canonical ensemble, implying that the confined fluid remains in chemical equilibrium with a bulk reservoir
of infinite extent. For each separation, the equilibrium (minimal) free energy is determined, and the resulting net pressure is evaluated either by
numerically differentiating the free energy or via the contact densities of monomers at the surfaces. All cluster configurations are included
following Boltzmann weighting, assuming that density variations occur solely in
the direction perpendicular to the walls \cite{Woodward:1991}.

In a completely ideal bulk cluster solution, where the only constraints are the intramolecular bonds and no intermolecular
interactions (point-like monomers) or external potentials are present, the free energy ${\cal F}_{clust}^{(id)}$ of the $r$-mers (clusters) of valency $Q$ can be
written exactly as
\begin{widetext}
\begin{equation}
\beta {\cal F}_{clust}^{(id)} = \int N_Q({\bf R})\left(\ln [N_Q({\bf R})] - 1\right)
d{\bf R} + \beta \int N_Q({\bf R}) V_b({\bf R}) d{\bf R} ,
\label{eq:idpol}
\end{equation}
\end{widetext}
where ${\bf R} = ({\bf r}_0,\dots,{\bf r}_{r-1})$ denotes a cluster configuration, and $N_Q({\bf R})$ is the distribution function such
that $N_Q({\bf R})d{\bf R}$ gives the number of clusters with configurations in the range $[{\bf R}, {\bf R}+d{\bf R}]$. $V_b({\bf R})$, the total
bond energy of a cluster with configuration ${\bf R}$, is evaluated by summing Eq.~\eqref{eq:hbond} over all its $r-1$ bonds.

For dissociated particles, the corresponding ideal free energy, ${\cal F}_{dis}^{(id)}$, is simpler:
\begin{equation}
  \beta {\cal F}_{dis}^{(id)} =  \sum_{\eta=\pm 1} \int n_d^\eta({\bf r}) \left(\ln [n_d^\eta({\bf r})] - 1 \right) d{\bf r}.
\end{equation}

The Helmholtz free energy for the bulk mixture can be expressed as
\begin{equation}
	\begin{split}
{ \cal F}(V,T,N_+,N_-,&n_d^+,n_d^-) = \sum_{Q=\pm1}{\cal F}_{clust}^{(id)}+ {\cal F}_{dis}^{(id)}\\
&+{\cal F}_{HS}^{(ex)}(n_m,n_d) + {\cal U}.
	\end{split}
\label{eq:helmholtz}
\end{equation}
Here, $N_+$ and $N_-$ distinguish between the configurational distributions of the two oppositely charged monovalent clusters, cationic and anionic, respectively.

In the bulk, both species contribute identically to the ideal free energy, while in confined or heterogeneous environments their free energies
differ. Excluded-volume effects are represented by ${\cal F}_{HS}^{(ex)}$, which depends on the total cluster monomer densities, $n_m$, as well
as the densities of the unconnected (dissociated) ions, $n_d$. This contribution is estimated using the Carnahan-Starling approximation\cite{Carnahan:69}, with
an adjustment factor of 0.8 to the total cluster member excluded volume, due to connectivity effects \cite{Woodward:92}. This is a rather crude estimate, but
as shown below, excluded volume only has a minor impact on the results reported in this work. All electrostatic terms are contained in the energy
term ${\cal U}$, and are treated within the mean-field approximation.

In the slit geometry, with the planar charged surfaces separated by $h$, we employ the grand potential $\Omega$, related to ${\cal F}$ by the Legendre transformation
\begin{widetext}
\begin{equation}
  \begin{split}
    \Omega(V,T,\mu_c,\mu_d,\Psi_D;&[N_+({\bf R'}),N_-({\bf R'}),n_d^+(z), n_d^-(z),\sigma_s])A^{-1}  =  \\       
      &    {\cal F}(V,T;[N_+({\bf R'}),N_-({\bf R'}),n_d^+(z),n_d^-(z),\sigma_s,V_{ex}(z,h)])A^{-1} \\
      &    -A^{-1}\int (\mu_c-e\Psi_D) N_+({\bf R'}) d{\bf R'} - A^{-1}\int (\mu_c+e\Psi_D) N_-({\bf R'}) d{\bf R'} \\      
      & - \int (\mu_d-e\Psi_D) n_d^+(z) dz - \int (\mu_d+e\Psi_D) n_d^-(z) dz - 2\pi k_BT l_B\sigma_s^2 h
\end{split}
\label{eq:grandpot}
\end{equation}
\end{widetext}
where $e$ is the elementary charge, $z$ denotes the coordinate perpendicular to the surfaces, ${\bf R'}$ represents the set of cluster coordinates that
are {\em constrained} to be within the slit regime (between the walls), $A$ is the surface area, and $\Psi_D$ is the Donnan potential that ensures
overall electroneutrality. The total Helmholtz free energy ${\cal F}$ accounts for contributions from configurational entropy, excluded-volume interactions, and
mean-field electrostatics, including the wall-wall interaction. It is minimised for canonical equilibrium density distributions, under the constraint of fixed
numbers of clusters and dissociated ions, under the influence of the wall exclusion potential $V_{ex}(z,h)$ and the surface charge
density $\sigma_s$ (at each wall).  The grand potential $\Omega$, on the other hand, is minimised for the equilibrium density distributions where the
clusters and simple ions are in chemical equilibrium with a bulk distribution with fixed (input) cluster salt and dissociated salt chemical potentials, $\mu_c$ and $\mu_d$.
The surface interaction potential $V_{ex}(z,h)$ imparts a non-electrostatic exclusion and is purely steric, being zero within the accessible
region ($d/2<z<(h-d/2)$) and infinite elsewhere. We arrive at the equilibrium grand potential, $\Omega_{eq}(h)$, by numerical minimisation, using
Picard iterations. Defining $g_s(h) \equiv \Omega_{eq}/A + p_b h$, where $p_b$ is the bulk osmotic pressure, the net interaction free energy is
given by $\Delta g_s \equiv g_s(h) - g_s(h_{max})$, with $h_{max} = 105$ {\AA}.

Upon solving for the equilibrium cluster ion densities, we need to perform repeated Boltzmann-weighted bond integrals, which effectively limits the
accessible cluster configurations. Moreover, the symmetry of the system will allow us to integrate away any dependence on $(x,y)$ coordinates. The
numerical procedure is probably best illustrated by examples. First, we define the following Boltzmann-weighted bond integral, $\Phi_b(|z|)$:
\begin{widetext}
\begin{equation}
\Phi_b(|z|) =
\begin{cases}
\displaystyle 2\pi\int_0^\infty \rho\,e^{-f(\rho^2+z^2)}\,d\rho
=\frac{\pi}{f}e^{-fz^2}, & |z|\ge d,\\[4mm]
\displaystyle 2\pi\int_{\sqrt{d^2-z^2}}^\infty \rho\,e^{-f(\rho^2+z^2)}\,d\rho
=\frac{\pi}{f}e^{-fd^2}, & |z|<d.
\end{cases}
\label{eq:phib}
\end{equation}
\end{widetext}

Given the symmetry of the system, and our mean-field assumption, which in turn is motivated by the presence of only monovalent species in a
high-dielectric solvent, we can write the expression for the equilibrium density of cationic central cluster ions, $n_{cent}^+(z)$, as:
\begin{equation}
\begin{split}
n_{cent}^+(z) = {}& e^{\beta\mu_c-\Lambda(z)-\beta e \Psi(z)} \\
&\times \left[\int \Phi_b(|z'-z|)\,e^{-\Lambda(z')-\beta e \Psi(z')}\,dz'
  \right]^{\frac{r-1}{2}} \\
&\times \left[\int \Phi_b(|z'-z|)\,e^{-\Lambda(z')+\beta e \Psi(z')}\,dz'
  \right]^{\frac{r-1}{2}} ,
\end{split}
\label{eq:ncent}
\end{equation}
where $\Lambda(z) \equiv \beta\, \delta{\cal F}_{HS}^{(ex)}(n_m,n_d)/\delta n_m$ and $\Psi(z)$ is the total electrostatic
potential at $z$ (including the Donnan potential). In order to arrive at the equilibrium satellite ion densities, it is useful
to first define two auxiliary functions $c_+(z)$ and $c_-(z)$:
\begin{equation}
\begin{split}
c_+(z) = {}& e^{-\Lambda(z)-\beta e \Psi(z)} \\
&\times \left[\int \Phi_b(|z'-z|)\,e^{-\Lambda(z')-\beta e \Psi(z')}\,dz'
  \right]^{\frac{r-1}{2}} \\
&\times \left[\int \Phi_b(|z'-z|)\,e^{-\Lambda(z')+\beta e \Psi(z')}\,dz'
  \right]^{\frac{r-1}{2}-1} ,
\end{split}
\label{eq:cplus}
\end{equation}
and
$c_-(z)= e^{2\beta e\Psi(z)}c_+(z)$.
They admit a
relatively compact expression for (say) the {\em total} equilibrium density of negative satellite ions, $n_{sat}^-(z)$, bonded to either a cationic or an anionic central ion:
\begin{equation}
\begin{split}
n_{sat}^-(z) = {}& \frac{r-1}{2}\,e^{\beta\mu_c-\Lambda(z)+\beta e \Psi(z)} \\
&\times \int \Phi_b(|z'-z|)(c_+(z')+c_-(z'))\,dz' ,
\end{split}
\label{eq:nsat}
\end{equation}
where the prefactor $(r-1)/2$ accounts for the equivalent anionic satellites. We hope that these examples suffice to illustrate how the
expressions for all equilibrium densities are obtained. In practice, these have to be iteratively solved, until self-consistent solutions
are obtained. We utilised a simple Picard scheme. Further details about general polymer-cDFT procedures are described in \cite{Nordholm:18}.

The inverse screening length, $1/\lambda$, at a given salt concentration is evaluated from the long-range
slope of $-\ln[p_{net}]$, where $p_{net} = -\partial g_s/\partial h$ is the net normal (\textit{disjoining}) pressure acting between
the surfaces. To facilitate direct comparison with SFA/AFM measurements, we invoke the Derjaguin approximation, $F(h)/R = 2\pi \Delta g_s(h)$, and present
our results as the force per radius, $F/R$, rather than the interaction free energy per unit area, $\Delta g_s(h)$.

\subsection{Monte Carlo simulations}
In order to highlight the extent to which the clusters modify the structure of the bulk fluid, Metropolis Monte Carlo (MC) simulations using the
star cluster model were also performed. By treating hard-sphere exclusion exactly and electrostatics without the mean-field approximation, these
simulations complement the cDFT treatment of star clusters in a slit geometry, which constitutes the main part of this work. Characterizing the
bulk structure is of independent interest, since clustering can only be inferred indirectly from surface force measurements, whereas the bulk
structure is in principle directly accessible to scattering experiments.

The simulations were carried out in the canonical ensemble, at $T = 298\,\mathrm{K}$, in a cubic box of side length $L$. Periodic boundary
conditions were applied
along all three Cartesian axes, thus simulating a bulk environment. 
All Coulomb interactions were evaluated using the tapered truncation, as suggested by Fanourgakis \cite{Fanourgakis:15}:
\begin{equation}
u(s) = \frac{1}{s} - \frac{7}{4R_t} + \frac{21 s^{4}}{4R_t^{5}}
- \frac{7 s^{5}}{R_t^{6}} + \frac{5 s^{6}}{2R_t^{7}} , \qquad s < R_t ,
\label{eq:taper}
\end{equation}
with $u(s) = 0$ for $s \geq R_t$, where $s$ denotes the separation between two ions and $R_t = L/2$ is the truncation radius. The non-bonded electrostatic
interaction between ions $i$ and $j$ then follows as $\beta\phi_{ij} = l_B\,\eta_i\eta_j\,u(s)$, in place of the Coulomb term of Eq. (\ref{eq:pairpot}), while
the hard-core term is retained. We have validated this treatment against computationally more expensive Ewald summations for closely related systems in
earlier work\cite{Forsman:24a,Ribar:24}, and we have reached the same conclusion for these systems (essentially identical results, also with simple
minimum-image truncation---not shown but data available upon request).

We implemented trial moves consisting of single-particle displacements of a cluster centre, of a satellite ion, or of a dissociated ion, selected at
random with fixed probabilities and accepted according to the standard Metropolis criterion. Satellites do not follow a displaced centre, so that
every bond of a cluster is re-evaluated whenever its central ion is moved, and cluster centre-of-mass motion is solely a consequence of accumulated single-particle displacements.

\begin{figure}
    \centering
    \includegraphics[width=0.97\linewidth]{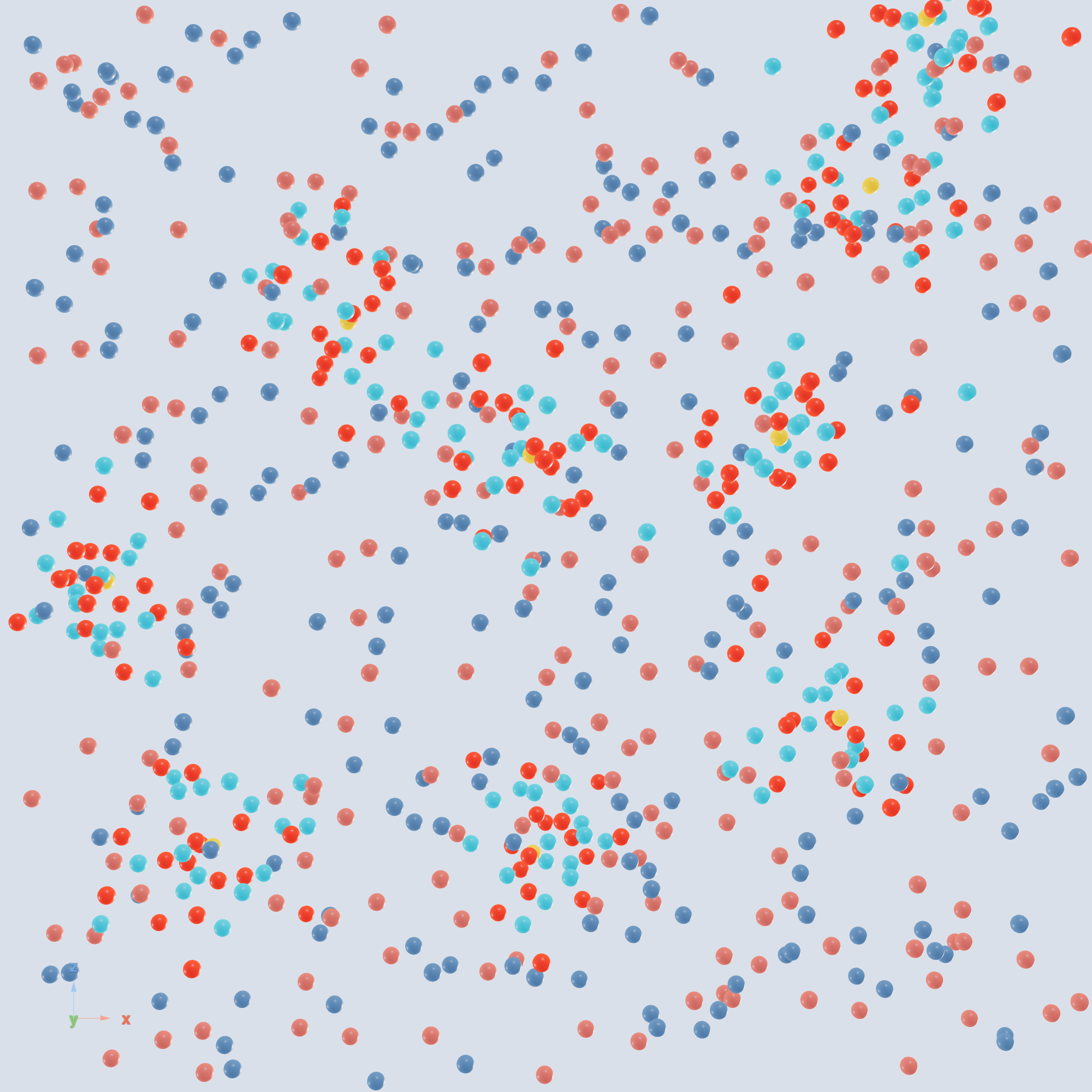}
    \caption{A configurational snapshot from a MC simulation of 33-mer star clusters at a total salt concentration
      of $c=2400\,\mathrm{mM}$, and $K^*=0.856$. The view shows
      all ions and clusters within a slab of 18 Å thickness. Yellow spheres represent cluster centres; red and cyan spheres are cationic and anionic satellites
      bonded to the centres; pale rose and slate spheres are dissociated free cations and anions, respectively. 
    }
    \label{fig:coords}
\end{figure}

An illustrative example of a final configurational snapshot with the harmonic bonds visible is presented in Figure \ref{fig:coords}.

\section{Results and Discussion}
\label{sec:results}

This work mainly focuses on cDFT predictions on how interactions between charged surfaces are affected if the salt solution displays partial
clustering, i.e., if the solution can be viewed as a mixture of star clusters and dissociated ions. Throughout, we distinguish between two contributions
to the effective screening length: an \textit{ionic} contribution, which accounts for how the total
charge is distributed between dissociated ions and clusters, and a \textit{polarisation} contribution, which arises from the
response of the clusters to the local electrostatic field. Nevertheless, it is also of interest to estimate bulk structural properties of such
model solution, since it might be possible to probe those structures experimentally, which would be an independent test of the clustering
hypothesis. These bulk structures will be analysed by canonical MC simulations, as described above.

\subsection{cDFT predictions: surface interactions and screening lengths}
Let us start by comparing surface forces in the presence and absence of
our model clusters at different overall (total) salt concentrations, $c$.

\begin{figure*}
    \centering
    \includegraphics{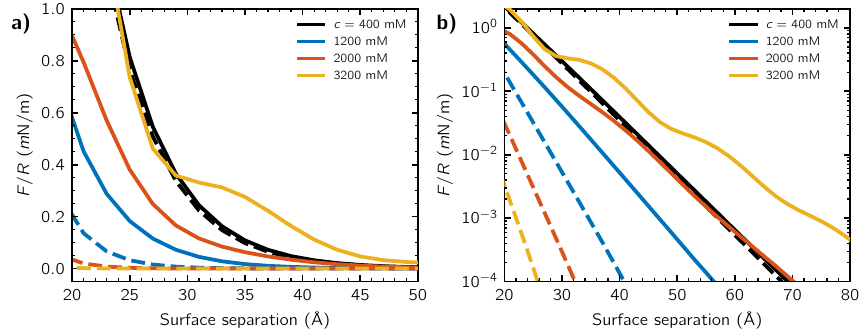}
    \caption{ \textbf{a)} Surface forces in the presence (solid) and absence (dashed) of star clusters. In the former case, the harmonic bond
      stiffness is set to $K^* = 0.856$, and the composition of the mixture at each overall salt concentration $c$ is specified in
      Table \ref{tab:cluster_concentrations}. \textbf{b)} presents the same data on a semilogarithmic plot, over a wider range of separations.}
       \label{fig:surfaceforces}
\end{figure*}

Such a comparison is presented in Figure \ref{fig:surfaceforces} for four representative total salt concentrations spanning the range
of Table \ref{tab:cluster_concentrations}.  As expected, the double layer repulsion decays monotonically for the model system composed
of 100\% dissociated ions, and is essentially of vanishing strength throughout the displayed separation regime, when $c>2000\,\mathrm{mM}$. In the
presence of our model clusters, however, an initial drop of the repulsion as more salt is added is followed by an \textit{increased} repulsion at
higher levels, i.e., the response is non-monotonic. The semilogarithmic representation in Fig. \ref{fig:surfaceforces}\textbf{b)} shows the slopes
of the surface forces steepening up to $1200\,\mathrm{mM}$ and flattening again thereafter, indicating an increase in the screening length rather than merely in the amplitude.

\begin{figure}[h]
\begin{center}
       \includegraphics[scale=1]{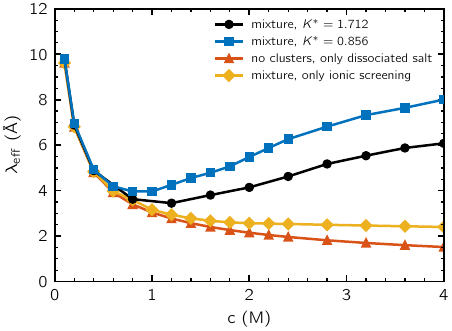}  
       \caption{The variation of the
         screening length, as obtained for various cDFT models. The mixture, ionic screening data points refer to a system where the concentration of the
         monovalent clusters is taken into account, but leaving out their polarisation contribution. 
       }
\label{fig:sclengths}
\end{center}
\end{figure}

In Figure \ref{fig:sclengths}, we demonstrate specifically, with calculations performed for all concentrations listed in Table \ref{tab:cluster_concentrations}, that
the screening length also displays a non-monotonic dependence. Note that solutions containing clusters are described as \textit{mixtures}. The minimum occurs at an
overall salt concentration close to $1\,\mathrm{M}$, with a weak dependence on the chosen harmonic force constant. These observations are in qualitative agreement
with most experiments that report \textit{anomalous underscreening}, although at least SFA measurements indicate a steeper rise in the anomalous regime. In
passing, we note that while we have assumed the clusters to be monovalent, essentially identical interactions are obtained with net-neutral clusters. This
is explicitly demonstrated in Appendix \ref{app:netcluster}.

A ``trivial'' aspect of clustering is that it renders the concentration of dissociated ions lower than the overall value, i.e., $c_d < c$. However, this
mechanism alone does \textit{not} suffice to explain the observed behaviour, since $c_d$ is still monotonically increasing as more salt is
added. The ``trivial'' effect is illustrated by the yellow line in Fig. \ref{fig:sclengths}; it corresponds to the
factor $\sqrt{C}$ of Eq. \eqref{eq:ratio}, which, since $\sqrt{C}\leq1$ by construction, can only shorten the decay relative to $\lambda_D$. It generates
a slower decay of the screening length with concentration, but never a non-monotonic dependence. In other words, the clusters themselves provide an
extra repulsive interaction between the surfaces.

So what is the origin of this additional repulsion that the clusters generate? One might be inclined to suggest that there are steric contributions
from the clusters, but these are actually quite small. We can demonstrate this by comparing our results with those obtained in the absence of excluded
volume, i.e., with a point-like representation of all ``spherical'' particles (${\cal F}_{HS}^{ex} = 0$).
\begin{figure}[h]
\begin{center}
       \includegraphics[scale=1]{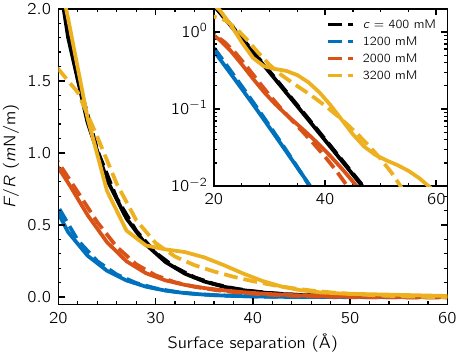}  
       \caption{\footnotesize{Surface forces in the presence of clusters, computed with and without excluded volume at $K^* = 0.856$. Solid lines are
           the full model, with a hard-sphere diameter $d = 3\,${\AA} for all species; dashed lines were obtained with point-like
           ions, $d = 0$. Concentrations are as in Fig. \ref{fig:surfaceforces}. The inset shows the same data on a semilogarithmic scale.
       }}
\label{fig:ideal}
\end{center}
\end{figure}
Such a comparison is made in Figure \ref{fig:ideal}. We see that, aside from a slight influence of an oscillatory contribution at very high ionic strengths
when $d=3$ Å, the results are almost identical. Since this contribution is absent when $d=0$ Å, we conclude it must arise due to excluded volume effects rather
than an oscillatory nature of the electrostatic decay. In fact, the minor difference observed indicates that the repulsion
is somewhat {\em stronger} in the absence of excluded volume interactions.

We argue that the origin of the cluster-induced repulsion is almost entirely due to polarisation, i.e., dielectric screening. Taking Eq.~\eqref{eq:ratio} into
account, the underscreening must originate in the remaining factor, $\sqrt{1+\chi_c}$, which is precisely the polarisability of the clusters themselves.

\subsection{Comparisons of cDFT calculations with the analytical model}

\begin{figure*}
    \centering
    \includegraphics[]{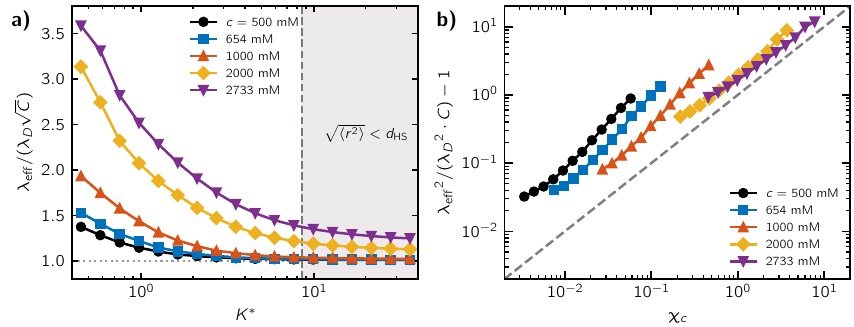}
    \caption{Effective screening lengths normalised according to Eq.~\eqref{eq:ratio}. All results are from cDFT, computed at varying total salt concentrations
      given in the legend. \textbf{a)} presents the normalised effective screening length at various harmonic bond stiffnesses, $K^*$. The grey area shows
      conditions where the root-mean-square bond length falls below the hard-sphere diameters. \textbf{b)} compares the agreement between the analytical
      theory and the observed response, Eq.~\eqref{eq:theory_prediction}; the dashed line is $x=y$.}
    \label{fig:master_eq_results}
\end{figure*}

With Figure \ref{fig:master_eq_results}, we demonstrate the qualitative agreement of the analytical theory with cDFT results. For each data-point, we computed
surface force calculations across the whole surface separation scan, and extracted the effective screening length from net pressure curves. Different colours
denote the different total concentrations of ions, with the cluster monomer concentration following Eq.~\eqref{eq:nm} and setting $r =21$ for all concentrations
as a fixed architecture to investigate the $K^*$ dependence. On Figure \ref{fig:master_eq_results}\textbf{a)} we identify that at each concentration
the softest bonds investigated produce the strongest screening enhancement, as predicted by theory. Secondly, we observe the strongest screening enhancement
at the highest total ion concentration and thus also highest cluster concentration. The grey area on the sub-plot denotes the region where the root-mean-square
bond length falls below the hard-sphere diameter of the ions, $\sqrt{\langle r^{2}\rangle}= l_B\sqrt{3/(2K^*)} < d = 3\,$ Å, i.e., $K^* \geq 8.5$. These data
points are included for completeness on sub-plot \textbf{a)}, but excluded from sub-plot \textbf{b)}. On sub-plot \ref{fig:master_eq_results}\textbf{b)} we
instead show the measured screening lengths as Eq.~\eqref{eq:theory_prediction}. The theoretical prediction should lie on the $x=y$ line (grey dashed line)
with no fitting parameters. We can observe qualitative agreement in the observed slopes at the high $\chi_c$ branch of each concentration series. This is
the soft-bond branch and it correctly predicts the strongest screening length enhancement. All series lie above the $x=y$ line, i.e., produce a response
which is stronger than analytically predicted. However, it is to be noted that the data points at higher total concentration, which contain correspondingly
more clusters, remain above the line but approach it increasingly closely. Two effects plausibly contribute to this non-ideal behaviour. Firstly, one has to
note that the analytical theory includes no intermolecular correlations, being built exclusively on the intramolecular charge structure of
point-charges. Secondly, the second-moment truncation underlying Eq.~\eqref{eq:theory_prediction} by construction removes non-vanishing contributions of higher order terms, an approximation that Eq.~\eqref{eq:exact_pole} lifts.  Nevertheless, the present theory qualitatively predicts a screening length
enhancement, which emerges exclusively from cluster polarisability, quantified by $\chi_c = \frac{1}{3}\kappa_{\rm sat}^2R_c^2$.

\subsection{Simulated bulk structure}
\begin{figure*}
\centering
       \includegraphics[scale=1]{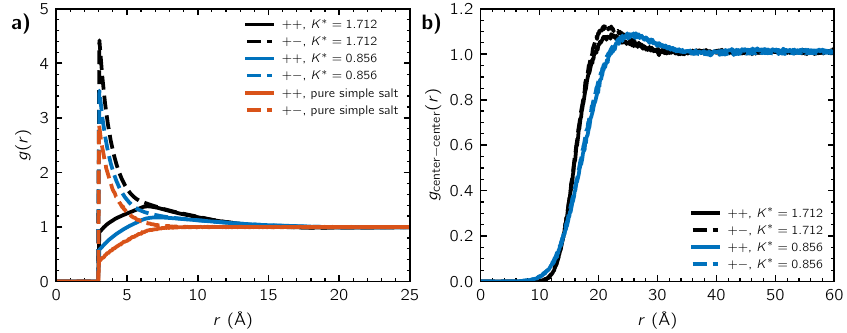}
       \caption{\textbf{a)} Simulated bulk radial distribution functions
           between {\em all} ions, i.e., with an indiscriminating approach. \textbf{b)} Simulated bulk radial distribution functions
           between cluster centers. The overall salt concentration for both sub-plots is $c=2400$mM.}
\label{fig:rdfall}

\end{figure*}

In Figure \ref{fig:rdfall}\textbf{a)}, we compare the simulated standard radial distribution functions (RDFs) obtained in a bulk of fully dissociated
ions, with those obtained from simulations of two different model cluster mixtures ($K^*=0.856$ and $K^*=1.712$), at a total salt concentration $c=2400\,\mathrm{mM}$.
In all three case we sampled distances between {\em all} ions, irrespective of their status (dissociated, satellite, central). In the presence of clusters, the
added intracluster attraction is qualitatively evident, but the relative difference is rather modest at close range. At intermediate separations, on the
other hand, we do observe some noticeable differences. With a system with only dissociated ions, the correlations have essentially vanished at separations
beyond about 10 {\AA} or so, but in the presence of clusters, a significant attractive correlation persists to roughly 15 {\AA}, for both investigated
bond strengths. In principle, such extended correlations might be detectable by scattering methods, though this conjecture remains to be verified. 

We can of course instead investigate correlations between cluster centres, as measured by their radial correlation function, $g_{center-center}(r)$. The results
are reported in Figure \ref{fig:rdfall}\textbf{b)}. As expected, we find a rather long-ranged ``hole'', beyond which there is a slightly attractive
correlation regime. Like- and opposite-charged cluster pairs give almost the same curve, indicating that these correlations are set by the size of the
clusters, rather than by their net charge, in line with Appendix~\ref{app:netcluster}, where the same conclusion was established from the surface
forces. An interesting finding is that these results are quite insensitive to the bond strength, at least within the investigated parameter regime. We
also note that a reduction of $f$ makes the clusters effectively larger, but also ``fluffier''. For the two studied bond strengths, the former effect
dominates, leading to an increased correlation hole as the bond strength drops, but in the limit of very weak bond strengths, the correlation hole will close up.

\section{Conclusion}
We have demonstrated that the formation of ion clusters at high concentrations in aqueous solutions does not only increase the {\em ionic} screening length
by depleting the population of dissociated ions. The ability of such clusters to polarise, in the vicinity of charged particles and surfaces, generates an
additional {\em dielectric} screening. It is this contribution, and not the depletion, that may lead to a non-monotonic dependence of the total effective
screening length, in qualitative agreement with the experimentally observed ``anomalous underscreening''. With the parameters that we have employed in
this work, the lowest screening length occurs at ionic strengths (approximately $1\,\mathrm{M}$) that agree well with experimental
observations. However, beyond this minimum, most SFA measurements report a stronger growth of the screening length than predicted by our numerical
calculations. Nevertheless, these results indicate that cluster polarisation is an important mechanism. One may even envisage deliberate synthesis
of ``ion-cluster-like'' molecules for applications where such stabilising mechanisms are useful.

\begin{acknowledgments}
  Jan Forsman acknowledges financial support by the Swedish Research Council, and
  computational resources from LUNARC. Last, but certainly not least, I (Jan Forsman) would
  like to extend my gratitude to 
  my colleague and friend, Rudi Podgornik. The first time I met Rudi, was when I
  was a PhD student, and he spent some months visiting our lab. He very generously shared
  his wisdom concerning theoretical approaches to polymer fluids and polyelectrolytes - concepts
  that proved useful to me in later research. Since then, we (and other co-workers) have collaborated on
  various topics, mainly
  involving electrostatic interactions in ionic fluids, where his ability to find solutions
  to difficult problems was indispensable.  I remember with particular fondness
  a research visit I made to Ljubljana. Not only did Rudi introduce me to a number
  of his colleagues, with fruitful discussions, he also gave me a guided tour 
  across Slovenia, including visits to caves, the (minute) coast line, and Ljubljana itself.
  It grieves me that I, due to temporary illness, was unable to attend the conference held in his honour. 
  Rudi will certainly be deeply missed, professionally as well as socially.
  It is an honour to dedicate this work to the memory of Rudolf Podgornik. 
\end{acknowledgments}

\section*{AUTHOR DECLARATIONS}

\subsection*{Conflict of Interest}

The authors have no conflicts to disclose.

\subsection*{Data availability statement}
The data that support the findings of this study are available from the corresponding author upon reasonable request.

\subsection*{Author Contributions}

\textbf{David Ribar}: Conceptualization (equal); Formal analysis (equal); Investigation (equal); Methodology (equal); Software (supporting);
Supervision (supporting); Validation (equal); Visualization (equal); Writing – original draft (equal); Writing - review \& editing (equal).
\textbf{Jake W. Felber}: Formal analysis (equal); Investigation (equal); Validation (equal); Visualization (equal); Writing – original
draft (equal); Writing - review \& editing (equal). \textbf{Clifford E. Woodward}: Validation (equal); Writing - review \& editing (equal).
\textbf{Jan Forsman}: Conceptualization (equal); Formal analysis (equal); Investigation (equal); Funding acquisition (lead); Methodology (equal);
Software (lead); Supervision (lead); Validation (equal); Visualization (equal); Writing – original draft (equal); Writing – review \& editing (equal).

\appendix

\section{Point-charge star-clusters: derivations}
\label{app:first}

This appendix collects the derivations underlying
Secs.~\ref{sec:theory:omega}--\ref{sec:theory:master}. For self-containment we
first restate the architecture. The central ion sits at the origin with index
$i=0$. The satellites, referred to as the \emph{corona}, carry indices $i=1,\dots,r-1$, where
$r$ is the total number of charges per cluster. Charge $i$ has valency
$\eta_i=\pm1$ at position $\mathbf{r}_i$ relative to the core, with
$\mathbf{r}_0=\mathbf{0}$ by construction. Each site carries a physical charge of $\eta_ie$. The corona is electroneutral,
$\sum_{i\ge1}\eta_i=0$, so the monovalent cluster carries the valency of its
core, $Q\equiv\eta_0+\sum_{i\ge1}\eta_i=\pm1$. All charges are points, the
satellites are statistically independent, and all bonds share the same
stiffness $f$. The intramolecular charge structure factor of a single cluster,
Eq.~\eqref{eq:omega_def}, is
\begin{equation}
    \omega_c(k) = \left\langle\left|\sum_{i=0}^{r-1}\eta_i
    e^{\mathrm{i}\mathbf{k}\cdot\mathbf{r}_i}\right|^{2}\right\rangle
    = \sum_{i=0}^{r-1}\sum_{j=0}^{r-1}\eta_i\eta_j
    \big\langle e^{\mathrm{i}\mathbf{k}\cdot(\mathbf{r}_i-\mathbf{r}_j)}\big\rangle ,
    \label{eq:omega_dsum}
\end{equation}
and the instantaneous reduced dipole moment is
$\mathbf{P}\equiv\sum_i\eta_i\mathbf{r}_i$, with the physical dipole moment $e\mathbf{P}$.

The harmonic bond $\beta V_b=f|\mathbf{r}'-\mathbf{r}|^{2}$ makes the satellite
position relative to the core Gaussian,
\begin{equation}
    p(\mathbf{r}) = (2\pi\xi^{2})^{-3/2}e^{-r^{2}/(2\xi^{2})},
    \qquad \xi^{2}\equiv\frac{1}{2f},
    \label{eq:pdist}
\end{equation}
with Fourier transform
\begin{equation}
    \tilde p(k) = \int_{\mathbb{R}^{3}}p(\mathbf{r})
    e^{-\mathrm{i}\mathbf{k}\cdot\mathbf{r}}\,\mathrm{d}\mathbf{r}
    = e^{-k^{2}\xi^{2}/2},
    \label{eq:ptilde}
\end{equation}
and mean-square bond length $\langle r^{2}\rangle=3\xi^{2}$. Expanding the exponential in Eq.~\eqref{eq:omega_dsum} with
$\mathbf{r}_{ij}\equiv\mathbf{r}_i-\mathbf{r}_j$, the zeroth-order term
factorises to $(\sum_i\eta_i)^{2}=Q^{2}$. The first-order term vanishes
configuration by configuration,
\begin{align*}
    \sum_{i,j}\eta_i\eta_j\langle\mathbf{r}_{ij}\rangle
    &= \Big(\sum_j\eta_j\Big)\Big(\sum_i\eta_i\mathbf{r}_i\Big)
     - \Big(\sum_i\eta_i\Big)\Big(\sum_j\eta_j\mathbf{r}_j\Big)\\
    &= Q\mathbf{P}-Q\mathbf{P}=0 ,
\end{align*}
since $\mathbf{r}_{ij}$ is antisymmetric under $i\leftrightarrow j$ while
$\eta_i\eta_j$ is symmetric. The same argument annihilates every odd order, so
the first neglected term is $\mathcal{O}(k^{4})$. At second order, fluid
isotropy permits $\langle(\mathbf{k}\cdot\mathbf{r}_{ij})^{2}\rangle
=k^{2}\langle|\mathbf{r}_{ij}|^{2}\rangle/3$, giving
\begin{equation}
    \omega_c(k) = Q^{2}-\frac{k^{2}}{6}\sum_{i,j}\eta_i\eta_j
    \big\langle|\mathbf{r}_i-\mathbf{r}_j|^{2}\big\rangle
    + \mathcal{O}(k^{4}),
    \label{eq:approx_omega}
\end{equation}
which with the definition of $R_c$,
\begin{equation}
	R_c^{2} \equiv \frac{3}{r-1}\lim_{k\to0}
	\frac{\omega_c(k)-Q^{2}}{k^{2}},
	\label{eq:appending_r_C}
\end{equation}
 is Eq.~\eqref{eq:omega_smallk} of the main text. Expanding $|\mathbf{r}_{ij}|^{2}=|\mathbf{r}_i|^{2}
-2\mathbf{r}_i\!\cdot\!\mathbf{r}_j+|\mathbf{r}_j|^{2}$ in
Eq.~\eqref{eq:approx_omega} and summing each piece,
\[
    \omega_c(k) = Q^{2}-\frac{k^{2}}{3}\left\{
    Q\sum_i\eta_i\big\langle|\mathbf{r}_i|^{2}\big\rangle
    - \big\langle|\mathbf{P}|^{2}\big\rangle\right\},
\]
so that
\begin{equation}
    (r-1)R_c^{2} = \big\langle|\mathbf{P}|^{2}\big\rangle
    - Q\sum_{i=0}^{r-1}\eta_i\big\langle|\mathbf{r}_i|^{2}\big\rangle ,
    \label{eq:C2split_app}
\end{equation}
which is Eq.~\eqref{eq:C2split}. Its second term is an origin-invariance
correction required only for net-charged clusters. For $Q=0$ the correction vanishes and
$R_c^{2}=\langle|\mathbf{P}|^{2}\rangle/(r-1)$.

Specialising to the harmonic star, the double sum
Eq.~\eqref{eq:omega_dsum} partitions into three groups. The $r$ diagonal terms
give $\sum_i\eta_i^{2}=r$. The core--satellite terms carry a single factor
$\tilde p(k)$ and vanish by corona electroneutrality. The off-diagonal
satellite--satellite terms factorise into $\tilde p(k)^{2}$ by satellite
independence, their charge sum reducing to $-(r-1)$. Hence
\begin{equation}
    \omega_c(k) = r - (r-1)e^{-k^{2}\xi^{2}},
    \label{eq:omega_star_app}
\end{equation}
which is Eq.~\eqref{eq:omega_star}. For this architecture the second term of
Eq.~\eqref{eq:C2split_app} vanishes,
\[
    \sum_{i=0}^{r-1}\eta_i\big\langle|\mathbf{r}_i|^{2}\big\rangle
    = \eta_0\big\langle|\mathbf{r}_0|^{2}\big\rangle
    + 3\xi^{2}\sum_{i=1}^{r-1}\eta_i = 0 ,
\]
the first term because the core sits at the origin, the second because uniform
stiffness allows $\langle|\mathbf{r}_i|^{2}\rangle=3\xi^{2}$ to be taken outside
the sum, whereupon corona electroneutrality applies. A charge-asymmetric corona breaks the electroneutrality step, while non-uniform
stiffness prevents $\langle|\mathbf{r}_i|^{2}\rangle$ from being taken outside the sum; and in either case Eq.~\eqref{eq:C2split_app} must be evaluated in full. Otherwise
$R_c^{2}=\langle|\mathbf{P}|^{2}\rangle/(r-1)$, and since the satellites are
independent with zero mean, $\langle|\mathbf{P}|^{2}\rangle=(r-1)\cdot3\xi^{2}$.
Expanding Eq.~\eqref{eq:omega_star_app} at small $k$ gives
$\omega_c-Q^{2}=(r-1)k^{2}\xi^{2}$ and hence the same result via
Eq.~\eqref{eq:appending_r_C}.

To demonstrate the general nature of this construction, we consider a rigid zwitterionic dimer of fixed bond length $d$
($r=2$, $Q=0$). For such a system Eq.~\eqref{eq:omega_dsum} contains two diagonal terms and two
cross terms,
\begin{equation}
    \omega_c(k) = 2 - 2\big\langle
    e^{\mathrm{i}\mathbf{k}\cdot\mathbf{r}_{01}}\big\rangle ,
    \label{eq:omega_dimer}
\end{equation}
and averaging the orientation of the rod uniformly over the sphere,
\begin{equation}
    \big\langle e^{\mathrm{i}\mathbf{k}\cdot\mathbf{r}_{01}}\big\rangle
    = \tfrac12\int_{-1}^{1}e^{\mathrm{i}kd\tau}\,\mathrm{d}\tau
    = \frac{\sin kd}{kd} ,
    \label{eq:sinc}
\end{equation}
gives $\omega_c(k)=2-2\,\mathrm{sinc}(kd)$ and, by
Eq.~\eqref{eq:rc_def}, $R_c^{2}=d^{2}$, with the polarisability now being
orientational rather than stretch-fluctuational. For such systems we identify
$\chi_c=\tfrac13\kappa_{\rm sat}^{2}d^{2}=\tfrac83\pi l_B n_c d^{2}
=\beta(2n_c)\mu^{2}/3\varepsilon_0\varepsilon_r$, with $\mu=ed$ the permanent
dipole moment of the pair and $2n_c$ the total dipole number density, which is
the known Debye--Langevin susceptibility of a dilute gas of rigid
dipoles~\cite{Frohlich1958}.

We turn now to the fluid response. An external test charge $q_0$ at the origin
sets up a potential $\psi$. A cluster responds as a unit: with its core at
$\mathbf{r}_0$ its Boltzmann weight is
$\exp[-\beta e\sum_i\eta_i\psi(\mathbf{r}_0+\mathbf{r}_i)]\simeq
1-\beta e\sum_i\eta_i\psi(\mathbf{r}_0+\mathbf{r}_i)$, so the density of cluster
\emph{centers} is modulated by
$\delta\rho_c(\mathbf{r}_0)=-n_c\beta e\sum_i\eta_i\psi(\mathbf{r}_0+\mathbf{r}_i)$,
and each such cluster deposits charge at all of its sites,
\[
    \delta\rho_{q,c}(\mathbf{r}) = \int\delta\rho_c(\mathbf{r}_0)
    \sum_{j=0}^{r-1}\eta_j e\,\delta(\mathbf{r}-\mathbf{r}_0-\mathbf{r}_j)\,
    \mathrm{d}\mathbf{r}_0.
\]
Index $i$ labels the site where the potential acted and $j$ the site where the
charge ended up, both on the same cluster. Fourier transforming and averaging
over internal coordinates therefore returns precisely
Eq.~\eqref{eq:omega_dsum}, giving
$\delta\tilde\rho_{q,c}(k)=-n_c\beta e^{2}\omega_c(k)\tilde\psi(k)$ per cluster
sign. Clusters of either sign contribute identically, since $\omega_c$ does not
depend on the sign of $Q$; adding these to the dissociated ions of both signs
gives $\delta\tilde\rho_q(k)=-\beta e^{2}\omega_{\rm tot}(k)\tilde\psi(k)$ with
$\omega_{\rm tot}$ as in Eq.~\eqref{eq:omegatot}. Poisson's equation
$\varepsilon_0\varepsilon_r k^{2}\tilde\psi=q_0+\delta\tilde\rho_q$ then yields
Eq.~\eqref{eq:psi}.

Inserting Eqs.~\eqref{eq:approx_omega} and \eqref{eq:appending_r_C} into
Eq.~\eqref{eq:omegatot}, the pole condition Eq.~\eqref{eq:pole} becomes
\[
    k^{2}\left\{1+\frac{8\pi l_B n_c}{3}(r-1)R_c^{2}\right\}
    = -8\pi l_B\left(n_d+n_cQ^{2}\right).
\]
The bracket is strictly greater than unity, so $k^{2}<0$ and the root must be purely
imaginary, $k\equiv \mathrm{i}\kappa_{\rm eff}$. It reduces to
\begin{equation}
    \kappa_{\rm eff}^{2} = \frac{8\pi l_B(n_d+n_cQ^2)}{1+\chi_c},
\end{equation} and referencing to $\lambda_D\equiv(8\pi l_B n_d)^{-1/2}$ gives
\begin{equation}
    \frac{\lambda_{\rm eff}}{\lambda_D} = \sqrt{1+\chi_c} \sqrt{C},
\end{equation}
with the concentration correction factor
\begin{equation}
    C \equiv \frac{n_d}{n_d+n_cQ^2}.
\end{equation}

\section{Net-neutral clusters}
\label{app:netcluster}
\begin{figure}[h!]
\begin{center}
       \includegraphics[scale=1]{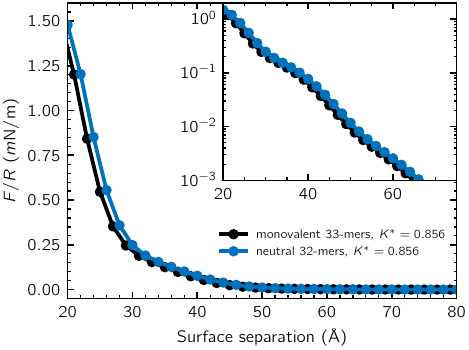}  
       \caption{Comparing cDFT results in a $2400\,\mathrm{mM}$ mixture, in the presence of neutral
           and monovalent clusters. 
       }
\label{fig:neut}
\end{center}
\end{figure}

To verify that the cluster net valency does not by itself drive the screening, we compare a monovalent 33-mer with a
net-neutral 32-mer at matched $K^*$ and matched total salt concentration in Fig.~\ref{fig:neut}. The two surface-force
curves are nearly indistinguishable, as anticipated by Eq.~\eqref{eq:master}. The polarisability response $\chi_c$ is
independent of $Q$ altogether, while the net valency enters only through $C$. Since the cluster net charges constitute
only about 2\% of the total ionic strength at this concentration, setting $Q=0$ lengthens the predicted decay by a mere
$1/\sqrt{C}-1 \approx 1\%$. The clusters therefore affect the decay length almost entirely through their polarisability,
rather than through the net-charge (monopole) channel.

\bibliography{poly}
\end{document}